\documentclass[useAMS,usegraphicx]{mn2e}

\title[Extragalactic MeV $\gamma$-ray emission from cocoons of young
radio galaxies]
{Extragalactic MeV $\gamma$-ray emission from cocoons of young
radio galaxies}
\author[M. Kino, N. Kawakatu, and H. Ito]
{M. Kino,$^{1}$ 
N. Kawakatu$^{2}$ and 
H. Ito$^{3}$ \\ 
$^{1}$ Department of Earth and Space Science, Osaka University, 
560-0043 Toyonaka, Japan\\
$^{2}$ National Astronomical Observatory of Japan, 181-8588 Mitaka, Japan\\
$^{3}$Department of Science $\&$ Engineering, Waseda University, 
Tokyo 169-8555, Japan}
\begin{document}

\date{}

\pagerange{\pageref{firstpage}--\pageref{lastpage}} \pubyear{2006}

\maketitle

\label{firstpage}

\begin{abstract}
Strong $\gamma$-ray emission 
from  cocoons of young radio galaxies is newly predicted.
Considering the process of adiabatic injection of 
the shock dissipation energy and mass of the relativistic jet 
in active nuclei (AGNs) into the cocoon, 
while assuming thermalizing electron plasma interactions,
we find that the thermal electron temperature of the cocoon is  
typically predicted in $\sim$MeV, which is 
determined only by the bulk Lorentz factor of the relativistic jet.
Together with the time-dependent dynamics
of the cocoon expansion,
we find that young cocoons can yield 
thermal bremsstrahlung emissions  at energies $\sim$ MeV.

\end{abstract}

\section{Introduction}

Relativistic jets in active galactic 
nuclei (AGNs) are widely believed to be 
the dissipation of kinetic energy of 
relativistic motion with a Lorentz
factor of order $\sim 10$ produced at the 
vicinity of a supermassive black hole at the
galactic center (Begelman, Blandford and Rees 1984 for reviews).
The jet in
powerful radio loud AGNs (i.e., FR II radio sources)
is slowed down via strong terminal shocks which are identified as 
hot spots.
The shocked plasma then expand sideways and 
envelope the whole jet system and this is so-called a cocoon
(Begelman and Cioffi 1989, hereafter BC89).
The cocoon is a by-product of the interaction
between AGN jets and surrounding intra-cluster medium (ICM).
The internal energy of the shocked plasma continuously 
inflates this cocoon. 
Initially,
the existence of the  cocoon is theoretically predicted 
by Scheuer (1974).

The first clear evidence for an
X-ray cavity was discovered
in the center of the Perseus cluster of galaxies
by  Boehringer et al. (1993).
The thermal ICM is displaced by the 
radio lobes which are composed of the remnants of the 
decelerated jet. Then the X-ray surface brightness
in those regions are significantly deceased.
These cavities correspond to the cocoons. 
Most of the X-ray cavities are 
associated with  low power AGN jets (i.e., FR I radio sources).
%
%
Recent X-ray observations of radio galaxies
shows us a further evidences of 
these X-ray cavities (e.g., Fabian et al. 2000; Blanton et al. 2001).
Another X-ray observational evidence of the cocoon 
is the non-thermal emission around radio lobes
(e.g., Feigelson et al. 1995; Isobe et al. 2002; Croston et al. 2005). 
In some cases, those non-thermal emissions are  associated with FR II
radio sources. 
%
%
In any case, there is no direct evidence of thermal emissions 
coming from the dilute thermal plasma inside the cocoon.

In this paper, 
we propose that ``a cocoon of a young radio galaxy''
as a new population of $\gamma$-ray emitters in the universe.
Up to now, little attention has been
paid to the evolution of 
thermal temperature and number density of the cocoon. 
Recently we have investigated 
the evolution of its temperature and number densities
by taking the proper account of mass and energy injections
by the relativistic jet 
(Kino and Kawakatu 2005; Kawakatu and Kino 2006).
We found that the cocoon remains  
constant temperature whilst the number density
increases as a cocoon becomes younger. 
This leads to our new prediction of
bright $\gamma$-ray emission from the young cocoon.

\section{ Cocoon inflation by dissipative relativistic jet}

Here we consider the time-evolution of expanding cocoon  
inflated by the dissipation energy of 
the relativistic jet via terminal shocks.
The adiabatic
energy injection into the cocoon is assumed here.
We will compare the source age $t$ and a cooling time scale 
and check the consistency at the last part of \S 2.
Note that the injection process of kinetic energy and mass
into the cocoon is ``continuous'' during $t$.
It is different from the  ``impulsive'' 
injection realized in gamma-ray bursts (GRBs)
and supernovae.

The time-averaged 
mass and energy injections from the  jet into
the cocoon, which govern 
the cocoon pressure $P_{\rm c}$ and 
mass density $\rho_{\rm c}$ 
are written as
\begin{eqnarray}\label{eq:pc}
\frac{{\hat \gamma}_{c}}{{\hat \gamma}_{c}-1}
\frac{P_{\rm c}(t)V_{c}(t)}{t}\approx
2 T^{01}_{\rm j}(t)  
A_{\rm j}(t) 
\end{eqnarray}
\begin{eqnarray}\label{eq:rho}
\frac{\rho_{\rm c}(t)V_{c}(t)}{t }\approx
2 J_{\rm j}(t)  
A_{\rm j}(t)   ,
\end{eqnarray}
where
${\hat \gamma}_{c}$,
$V_{\rm c}$,
$T^{01}_{\rm j}$, 
$J_{\rm j}$ and
$A_{\rm j}$,
are
the adiabatic index of the plasma in the cocoon,
the volume of the cocoon,
the kinetic energy and mass flux of the jet, and 
the cross-sectional area of the jet,
respectively.
%
The total kinetic energy and mass flux of the jet are
$T_{\rm j}^{01}=\rho_{\rm j}c^{2}\Gamma_{\rm j}^{2}v_{\rm j}$,
$J_{\rm j}=\rho_{\rm j}\Gamma_{\rm j}v_{\rm j}$ where
$\rho_{\rm j}$, and 
$\Gamma_{\rm j}$ are
mass density and bulk Lorentz factor of the jet
(Blandford and Rees 1974). 
Hereafter we set $v_{\rm j}=c$.
The total kinetic power of the relativistic jet 
is defined as 
$L_{\rm j}\equiv 2 T^{01}_{\rm j}(t) A_{\rm j}(t)$ 
and it is assumed to be constant in time.
Although little attention has been payed to the
mass injection Eq. (\ref{eq:rho}) up to now,
it is of great significance 
to take account of the Eq. (\ref{eq:rho})
for deriving the cocoon temperatures.
Hence we take up the the Eq. (\ref{eq:rho})
to evaluate the temperatures of the cocoons.
In contrast, the energy equation 
Eq. (\ref{eq:pc}) has been widely utilized 
in the literatures of 
the AGN bubbles in  various ways 
(e.g., BC89; Dunn and Fabian 2004).

The kinetic power dominance in the 
flow is postulated in this work
in accordance with the observational indications  
(e.g., Leahy and Gizani 2001; Isobe et al. 2002; Croston et al. 2005). 
The jet is assumed to be cold since
the hot plasma produced at the central engine usually cool down very 
quickly (e.g., Iwamoto and Takahara 2004).
As for the mass and kinetic energy flux of 
powerful relativistic jets,
numerical simulations tell us that
no significant entrainment of the environmental matter
takes place during the jet propagation (Scheck et al. 2002).
According to this, 
the mass and kinetic energy flux of the jet are 
regarded as constant in time.
Then, the conditions of $T_{\rm j}^{01}={\rm const.}$, and
$J_{\rm j}={\rm const.}$ leads to the important relations of  
\begin{eqnarray}
\rho_{\rm j}(t)A_{\rm j}(t)={\rm const}, \quad
\Gamma_{\rm j}(t)={\rm const}.
\end{eqnarray}
In fact, the constant $\Gamma_{\rm j}$ 
agrees with the relativistic hydrodynamic simulations
(e.g., Marti et al. 1997; Scheck et al. 2002).
In order to evaluate $L_{\rm j}$,  we use
the shock jump condition of 
$\Gamma_{\rm j}^{2}\rho_{\rm j} =\beta_{\rm hs}^{2}
\rho_{\rm ICM}$ (Kawakatu and Kino 2006)
where 
$\beta_{\rm hs}(=v_{\rm hs}/c)$ and
$\rho_{\rm ICM}$ is the  
advance speed of the  hot spot $\beta_{\rm hs}= 10^{-2}\beta_{-2}$ 
(Liu et al. 1992; Scheuer 1995) and
the mass density of ICM, respectively.
Using, the jump condition, 
$L_{\rm j}$ is given by 
\begin{eqnarray}
L_{\rm j}
=2 \times 10^{45}~
R_{\rm kpc}^{2}
\beta_{-2}^{2}
n_{-2}
~{\rm erg \ s^{-1}}
\end{eqnarray}
where
we use $A_{\rm j}(t)=\pi R_{\rm hs}^{2}(t)$, and 
the hot spot radius $R_{\rm hs}$ is given by
$R_{\rm kpc}=R_{\rm hs}(10^{7}~{\rm yr})/1~{\rm kpc}$.
As a fiducial case, we set
the number density of surrounding ICM as  
$n_{\rm ICM}(d)=\rho_{\rm ICM}(d)/m_{p}= 10^{-2}~{\rm cm^{-3}} 
n_{-2} (d/30~{\rm kpc})^{-2}$ 
where $d$ is the distance from
the center of ICM and cocoon.
Since the change of the index from $-2$
does not change the essential 
physics discussed in this work,
we focus on this case for simplicity.
Since $L_{\rm j}$ is the ultimate source of the
phenomena associated with the cocoon, all of the 
emission powers which will appear in \S 3 
should be less than $L_{\rm j}$.

The number density of total electrons in 
the cocoon $n_{e}(t)$ is 
governed by the cocoon geometry and 
its plasma content.
For convenience, we define
the ratio of 
``the volume swept by the unshocked relativistic jet''
to ``the volume of the cocoon''
as ${\cal A}(t)$.
Hereafter we denote 
$V_{\rm c}(t)=2(\pi/3){\cal R}^{2}Z_{\rm hs}^{3}(t)$, 
$Z_{\rm hs}$ satisfies $Z_{\rm hs}(t)=\beta_{\rm hs}ct$, 
$R_{\rm c}$, and 
${\cal R}\equiv R_{\rm c}/Z_{\rm hs}<1$  as
the cocoon volume,
the distance from the central engine to the hot spot, 
is the radius of the cocoon body, and 
the aspect-ratio of the cocoon, respectively
(e.g., BC89; Kino and Kawakatu 2005).
Postulating that ${\cal R}$ and 
$Z_{\rm hs}/R_{\rm hs}$ are constant in time, 
${\cal A}(t)$ is evaluated as
\begin{eqnarray}\label{eq:A}
{\cal A}(t)\equiv
\frac{2A_{\rm j}(t)v_{\rm j} t}{V_{\rm c}(t)}
\approx
0.4~
{\cal R}^{-2}
R_{\rm kpc}^{2}
Z_{30}^{-2}
\beta_{-2}^{-1} ,
\end{eqnarray}
where
$Z_{30}= Z_{\rm hs}(10^{7}~{\rm yr})/30~{\rm kpc}$.
Note that, in the case,
the time dependence of ${\cal A}$ is deleted
since $V_{\rm c}\propto t^{-3}$ and $A_{\rm j}\propto t^{2}$.
We stress that 
this case satisfies 
the observational indication of $v_{\rm hs}={\rm const}$
(e.g., Conway 2002).
Eq. (\ref{eq:A}) tells us that ${\cal A}$ is of order unity.
Actually it is  seen in some numerical simulations
(e.g., in Fig. 2 of Scheck et al. 2002).
The cocoon mass density $\rho_{c}(t)$
is controlled by the mass injection by the jet and 
it can be expressed as
\begin{eqnarray} 
\rho_{\rm c}(t)
&\approx&
\Gamma_{\rm j}
\rho_{\rm j}(t)
{\cal A}  \nonumber \\
&=& 
\beta_{\rm hs}^{2}
\Gamma_{\rm j}^{-1}
\rho_{\rm ICM}(Z_{\rm hs}(t))
{\cal A}  ,
\end{eqnarray}
where we use the shock condition of 
$\Gamma_{\rm j}^{2}\rho_{\rm j} =\beta_{\rm hs}^{2}
\rho_{\rm ICM}$. 
Adopting typical quantities of FR II sources
(Begelman, Blandford and Rees 1984;
Miley 1980; Bridle and Perley 1984),
we obtain
\begin{eqnarray} \label{eq:ne}
n_{e}(t)
\approx
4 \times 10^{-5} 
\bar {{\cal A}}
n_{-2}
\Gamma_{10}
\beta_{-2}^{2}
\left(\frac{t}{10^{7}~{\rm yr}}
\right)^{-2}
{\rm cm^{-3}}
\end{eqnarray}
where
$\Gamma=10 \Gamma_{10}$, and
$\bar{{\cal A}}={\cal A}/0.4$. 
Here
we assume that the mass density of 
the $e^{\pm}$ pair plasma is heavier
than that of electron-proton one,
and then we adopt  $\rho_{\rm c}\approx 2 m_{e}n_{e}$
in the light of previous works 
(Reynolds et al. 1996; 
Wardle et al. 1998;
Sikora and Madejski 2000;
Kino and Takahara 2004). 
However, the mixture ratio of $e^{\pm}$ pair
and electron-proton is still open.
If we assume completely 
pure electron-proton content in the jet,
too small $n_{e}$ is required and it
conflict with that of non-thermal electrons
(Kino and Takahara 2004).

%

Let us estimate
the electron (and positron) temperature ($T_{e}$)
and proton temperature ($T_{p}$).
From Eqs. (\ref{eq:pc}) and (\ref{eq:rho})
together with the equation of state 
\begin{eqnarray}
P_{\rm c}\approx  2n_{e}kT_{e} ,
\end{eqnarray}
we can directly derive the temperatures as
\begin{eqnarray} \label{eq:Te}
kT_{e} \approx 1~ \Gamma_{10} ~ {\rm MeV}, \quad
kT_{p} \approx 2 ~ \Gamma_{10} ~ {\rm GeV}
\end{eqnarray}
where we adopt the two temperatures condition of
$kT_{e}\approx (m_{e}/m_{p}) k T_{p}$.
It should be stressed that 
the temperatures are governed only by $\Gamma_{\rm j}$.
It is  also worth noting that the geometrical
factors in Eqs. (\ref{eq:pc}) and (\ref{eq:rho}) are completely
cancelled out.
Actually, the $\Gamma_{\rm j}$ dependence
of Eq. (\ref{eq:Te}) well coincide with 
the result of  hydrodynamic simulations of 
relativistic outflows 
(Fig. 5 in Mart\'{i} et al. 1997).
One can naturally understand these properties
by comparing the well-established properties 
such as supernovae and GRBs.
Constant temperature in AGN jet can be realized by 
the ``continuous'' energy 
injection into the expanding cocoon 
whilst temperatures of 
astrophysical explosive sources such as gamma-ray bursts
and supernovae would be decreased
because of ``impulsive'' injection of the energy. 
Since the shock dissipation 
of the relativistic flow into non-relativistic one, in general,
requires the energy conversion of  whole kinetic energy density
$\Gamma_{\rm j}\rho_{\rm j}c^{2}$ into internal one
(Piran 1999). 
Thus the resultant temperatures are
uniquely governed by $\Gamma_{\rm j}$ and they remain to be 
constant in time. 
Similarly, in the studies of continous steller winds,
the constant temperature has been predicted for
a hot interior consist of the shocked wind (Weaver et al. 1977).

Here we examine
the time scale of the Coulomb interaction  
between protons and electrons.
The time-scale of energy transfer from 
the protons to electrons is given by
$t_{ep} \approx(n_{p}\sigma_{t}c)^{-1}$
where 
$\sigma_{t}=4\pi (e^{2}/kT_{e})^{2} \ln \Lambda_{c}$
is the transport cross section
for electron-proton collision.
The coulomb logarithm is written as
$\ln\Lambda_{c}\approx \ln(3 kT_{e}\lambda_{D}/e^{2})$ 
where 
$\lambda_{D}=(kT_{e}/4\pi n_{e}e^{2})^{1/2}$
is the Debye length (Totani 1998). 
A typical case of hot spots in AGN jets,
we obtain $\ln \Lambda_{c}\sim 50$. 
Therefore, even using the maximal
proton number density $n_{p}\approx n_{e}$,
$t_{ep}$ satisfies 
\begin{eqnarray} \label{eq:ep}
\frac{t_{ep}(t)}{t}\sim
5\times 10^{3} ~
\Theta_{10}^{2}
\bar{n}_{e}^{-1}
\left(\frac{t}{10^{7}~{\rm yr}}\right)
\end{eqnarray}
where 
$\Theta_{e}\equiv kT_{e}/m_{e}c^{2} = 10\Theta_{10}$, and 
$\bar{n}_{e}
\equiv n_{e}(10^{7}~{\rm yr})/4\times10^{-5}~{\rm cm^{-3}}$,
are
the electron temperature in unit of $m_{e}c^{2}$, and
normalized number density of thermal electrons,
respectively.
As mentioned before, recent studies
suggest the existence of large amount of 
$e^{\pm}$ pairs in  AGN outflows
which lead to much smaller $n_{p}$.
Hence Eq. (\ref{eq:ep}) shows the 
minimum value of $t_{ep}(t)/t$.
Thus 
the energy transfer from protons to electrons is inefficient
by the Coulomb coupling unless the cocoon is much younger than 
$t\sim 10^{4}~{\rm yr}$.

Next we evaluate the time scale of 
thermal bremsstrahlung cooling in the cocoons.
It is well known that thermal bremsstrahlung is 
inefficient for the dilute plasma since its emissivity
shows $\propto n_{e}^{2}(t)$ where $n_{e}(t)$
is the 
electron number density in the cocoon.
For the shock-heated 
electrons with the temperature of 
$\Theta_{e}\approx \Gamma_{\rm j}$,
the cooling time of 
the bremsstrahlung per unit volume 
is estimated as
$t_{\rm brem}\approx
\Gamma m_{e}c^{2}n_{e}/\epsilon_{\rm brem}$
where 
the bremsstrahlung emissivity  
in the relativistic regime is
$\epsilon_{\rm brem}=1.3\times 10^{-22} 
\Theta_{e}^{1/2}n_{e}^{2}(1+2.6\Theta_{e})~{\rm erg~ cm^{-3}~s^{-1}}$
with the Gaunt factor of 1.2 (Rybicki and Lightman 1979).
The condition of $t_{\rm brem}(t)>t$ 
\begin{eqnarray} \label{eq:brem}
\frac{t_{\rm brem}(t)}{t}\approx
5\times 10^{4} ~
\Theta_{10}^{-1/2}
\bar{n}_{e}^{-1}\left(\frac{t}{10^{7}~{\rm yr}}\right)
\end{eqnarray}
actually  holds.
Therefore most of the shock dissipation energy is deposited into
the cocoon without suffering strong radiative cooling
and our treatment of adiabatic energy injection 
in Eq. (\ref{eq:pc}) is verified
for $t< t_{\rm brem}(t)$. 
We limit our attention on this case
in the present work.

On the thermalizations of electrons and protons,
it is worth to refer the recent studies
with the particle-in-cell (PIC) simulations,
since we only examine the simple case of 
the classical Coulomb interaction.
PIC simulations begin to shed light on the complicated 
microscopic dynamics with in the 
relativistic collisionless shock.
Using one-dimensional PIC simulations, 
Shimada and Hoshino (2000) revealed that
the collision and merging processes among the 
coherent waves are accompanied by the 
strong thermalization of electrons.
The results of three-dimensional PIC simulations 
(Nishikawa et al. 2003; Frederiksen et al. 2004) 
also tell us that electron populations are 
quickly thermalized whilst ion
population tend to retain distinct bulk speed and 
thermalize slowly.
In the result of Shimada and Hoshino (2000),
(Fig. 4 in their paper), 
we see that the proton energy is transferred
to the electrons, then the electrons are heated-up by protons.
To sum up, PIC simulations imply the
quick thermalization of electron populations 
by plasma waves and their associated
instabilities such as the two-stream instability.
Therefore, our estimate of $T_{e}$ would correspond to
the lower limit of $T_{e}$.
It is not the purpose of this paper to 
derive more realistic $T_{e}$ in detail.

\section{Emissions from a young cocoon}
\subsection{Thermal MeV bremsstrahlung emission}

The time-dependence of 
the thermal bremsstrahlung luminosity
$L_{\rm brem}$ is given by $L_{\rm brem}(t)\propto
n_{e}^{2}(t)T_{e}^{3/2}V_{\rm c}(t)
\propto t^{-1}$ based on the cocoon expansion
shown in the previous section.
Hence it is clear that a younger cocoons are 
brighter bremsstrahlung emitters than older cocoons.
In a similar way,
brighter synchrotron luminosity has been expected
for younger radio galaxies (Readhead et al. 1996; Begelman 1996).
With relativistic thermal
bremsstrahlung emissivity (Rybicki and Lightman 1979), 
the luminosity of the 
optically thin thermal bremsstrahlung emission
$\nu L_{\nu}$ at energies $\sim 1~{\rm MeV}$ 
is estimated as 
\begin{eqnarray}\label{eq:l_brem}
L_{\rm brem}(t)\approx 2\times 10^{40}~
\bar{n}_{e}^{2}
{\cal R}^{2}
\Theta_{10}^{3/2}
\left(\frac{t}{10^{7}~{\rm yr}}\right)^{-1}~{\rm erg \ s^{-1}} .
\end{eqnarray}
Here we omit the redshift ($z$) factor  
merely for simplicity.
Eq.  (\ref{eq:l_brem}) explains the reasons why for 
the no detection of the thermal emission from  
older cocoons.
One is simply because it is not very bright.
The other is because the predicted energy range is $\sim 1~{\rm MeV}$,
the MeV-$\gamma$ astronomy is still immature and it is sometimes 
called as ``sensitivity gap'' compared with 
the energy range below 10 keV 
and above GeV ranges (Takahashi et al. 2004).

For example,
the bremsstrahlung emission from the cocoon
located at a typical distance of $D=10^{3}~{\rm Mpc}$
(O'Dea and Baum 1998) is examined here. 
In Fig. 1, we show 
the predicted values of $\nu F_{\nu}$
for the cocoons with $t=10^{7}~{\rm yr}$ and 
$t=10^{4}~{\rm yr}$. 
The cocoon with $t=10^{7}~{\rm yr}$
have $\nu F_{\nu}\sim 10^{-14}~{\rm erg~cm^{-2}s^{-1}}$.
The detection threshold of 
SPI instrument on board the INTEGRAL satellite is about
$\nu F_{\nu}\sim 10^{-9}~{\rm erg~cm^{-2}s^{-1}}$ 
at $\sim 1~{\rm MeV}$.
For a young cocoon with $t=10^{4}~{\rm yr}$, the 
predicted luminosity is 
$\sim 10^{3}$ times larger than that 
$\nu F_{\nu}\sim 10^{-11}~{\rm erg~cm^{-2}s^{-1}}$. 
This is still less than the threshold of 
INTEGRAL. This may be the reason why no clear
detection of MeV emission from young cocoons up to now.
Fig. 1 shows that 
the XMM/Newton satellites can detect the 
low energy part of the thermal bremsstrahlung from young cocoons
in principle. 
Hence some of extragalactic unidentified X-ray sources
could be attributed as
the low energy tail of the bremsstrahlung emissions. 
Interestingly, 
the recent observation by XMM/Newton reveals that 
the spectrum of young radio-loud AGN B1358+624
actually shows the power-law slope close
to the bremsstrahlung's one (Vink at al. 2006).

In MeV energy band, a proposed mission of detector
SGD on board the NeXT satellite 
with the eye up to $\sim 0.6~{\rm MeV}$ (Takahashi et al. 2004)
could detect the thermal MeV emission 
from those located slightly 
closer and/or younger with smaller Lorentz factor.
Lastly,
other future mission, Advanced Compton Telescope (ACT),
is worth to be noticed.
The sensitivity of ACT is expected to be a significantly 
improved compared with INTEGRAL one 
(http://heseweb.nrl.navy.mil/gamma/detector/3C
/3C$\_$sens.htm).
Although 
they focus on the super-nova science (Milne et al. 2002)
for the moment, the ACT would be 
a promising tool also for the young cocoon science.

At a glance, one may think it hard to distinguish 
overlapping emissions from the core of the AGN
with limited spacial angular resolution
of current satellites.
Time variabilities of 
observed spectra is the key to distinguish them. 
It is obvious that the cocoon emission is 
constant in time whilst various emissions
from the core of AGN should be highly variable.
Hence steady emissions are
convincingly originated in cocoons.
Furthermore, the averaged spectral index of 
AGN core emissions at X-ray band 
(Koratkar and Blaes 1999; Kawaguchi et al. 2001) are softer
than the bremsstrahlung emission discussed in the present work. 
Hence the difference of the spectral index is also a 
useful tool to figure out the origin of the emission.

\begin{figure}
\includegraphics[width=8cm]{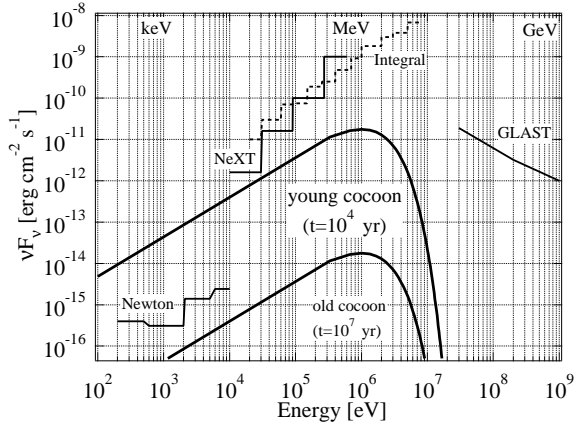}
\caption
{Model prediction of MeV-peaked thermal bremsstrahlung
emission from cocoons located at $D=10^{3}$ Mpc. 
The predicted emission from 
young cocoon is brighter enough to detect in X-ray band
whilst that from an old cocoon is much darker than 
the detection limits 
(Hasinger et al. 2001; Roques et al. 2003; Takahashi et al. 2004).}
\label{fig:MeV}
\end{figure}

\subsection{Non-thermal emissions}

Non-thermal emission from AGN jets is 
another key ingredient to investigate their physics. 
Non-thermal synchrotron emission from the radio lobes 
due to the relativistic electrons is well known
characteristic of AGN jets 
(Miley 1980; Bridle and Perley 1984).
Recently, inverse-Compton (IC) emissions from 
large scale jets have been also intensively 
explored both theoretically and observationally
 (e.g., Celotti and Fabian 2004; Croston et al. 2005).

The properties of 
IC emissions from young cocoons is discussed here.
For this purpose,
we firstly consider the properties 
of the synchrotron emission.
The magnetic flux conservation is assumed here
during the jet propagation which is given by
\begin{eqnarray}
B_{\rm hs}(t)R_{\rm hs}^{Y}(t)=const. \quad (1\le Y \le 2) \nonumber
\end{eqnarray}
where $Y$ is a parameter expressing the configuration 
of the magnetic field in the hot spot.
The magnetic flux from the central 
engine is assumed to be constant in time.
The case of constant $Y=1$ shows  
the purely toroidal-dominated magnetic field 
whilst $Y=2$ is relevant to the purely 
poloidal-dominated magnetic field.
Using $Y$,
the time dependence
of the synchrotron luminosity at the hot spot 
$L_{\rm hs,syn}(t)$ may be given by 
$L_{\rm hs,syn}(t)
\propto R_{\rm hs}^3(t)\gamma^{2}
B_{\rm hs}^{2}(t)n_{e}^{\rm NT}(\gamma,t)
\propto t^{-2Y+1}$
where 
we assume that the number density 
of the non-thermal electrons $n_{e}^{\rm NT}(\gamma,t)$
is proportional to $n_{e}(t)$, 
and $\gamma$ is constant in time because
the synchrotron cooling time 
tend to be longer than the sound crossing time 
at the hot spot (e.g., KT04).
Taking the observational fact of the 
large number of the CSOs in spite of their young age,
the larger synchrotron luminosity is required
for younger sources (Begelman 1996; Readhead at al. 1996).
Qualitatively, the  model 
well reproduce this observational properties
of young radio galaxies.

To evaluate IC emission of the cocoon,
it may be useful to define the quantities of  
$f_{\rm ssc}(t)\equiv U_{\rm ssc}(t)/U_{\rm syn}(t)$ and 
$f_{\rm IC/CMB}(t)\equiv U_{\rm IC/CMB}(t)/U_{\rm syn}(t)$,
where
$U_{\rm syn}$,
$U_{\rm ssc}$, and
$U_{\rm IC/CMB}$, are
the energy densities of 
synchrotron photons,
those of synchrotron-self Compton (SSC), 
and those of IC scattering of the 
Cosmic-Microwave Background (CMB),
respectively.
Photons with larger density is the 
dominant seed photons for IC scattering. 
We denote that the IC luminosities for 
synchrotron photons and the CMB as
$L_{\rm ssc}$ and $L_{\rm IC/CMB}$, respectively. 
%

%
For $U_{\rm ssc}(t)< U_{\rm CMB}$,
we see that $f_{\rm IC/CMB}(t)\propto t^{2Y}$ 
which implies that younger cocoon produce less IC/CMB photons
in contrast to the case of bremsstrahlung ones.
It is of great importance to examine
whether the predicted frequency of 
IC/CMB emission overlapping in MeV band or not.
According to the standard diffusive
shock acceleration, the acceleration time scale
is estimated as 
$t_{\rm acc}=(2\pi \gamma m_{e}c\xi)/(eB_{\rm hs})$
where $\xi$ is the parameter characterizing the
mean free path for the scattering (e.g., Drury 1983).
The maximum Lorentz factor of electrons $\gamma$
can be obtained 
by the equating $t_{\rm acc}$ to the
synchrotron cooling time 
$t_{\rm syn}=(6\pi m_{e}c^{2})
/(\sigma_{\rm T}\gamma c B_{\rm hs}^{2})$.
This shows the $B_{\rm hs}$ dependence of 
$\gamma$ as
\begin{eqnarray}\label{eq:gammaB}
\gamma(t)\propto B_{\rm hs}^{-1/2}(t)   .
\end{eqnarray}
Assuming that 
the strength of magnetic field
in the cocoon $B(t)$ 
is proportional to the one in the hot spot, 
the maximum frequency of the IC/ICM emission
$\nu_{\rm IC/CMB}
\propto \gamma^{2}\nu_{\rm CMB}$ can be estimated as
\begin{eqnarray}
\nu_{\rm IC/CMB}(t) \sim 
1\times 10^{19}\gamma_{4}^{2}
\left(
\frac{t}{10^{7}~{\rm yr}}
\right)^{Y}~{\rm Hz}  ,
\end{eqnarray}
where we denote
$B(t)\propto B_{\rm hs}(t)\propto R_{\rm hs}^{-Y}(t)\propto t^{-Y}$,
$\gamma(t)=10^{4}\gamma_{4} (t/10^{7}~{\rm yr})^{Y/2}$, and 
$\gamma_{4}=\gamma/10^{4}$.
The typical value of $\gamma$ is 
adopted from Blandford (1990).
From this, one can find that $\nu_{\rm IC/CMB}(t)$
of young cocoons much smaller than
$1~{\rm MeV}=2\times 10^{20}~{\rm Hz}$. 


In the case of  $U_{\rm syn}(t) > U_{\rm CMB}$, 
the behaviour of $U_{\rm syn}(t)$ is given by 
$L_{\rm syn}(t)\propto c Z_{\rm hs}^{2}(t)U_{\rm syn}(t)
\propto \epsilon_{\rm syn}(t)V_{\rm c}(t)$.
From this we obtain $U_{\rm syn}(t)\propto U_{\rm B}(t)t^{-1}$.
The model predicts that $f_{\rm ssc}(t)\propto t^{-1}$ in time 
and younger cocoon yields more SSC photons.
Using this, $L_{\rm ssc}$  can be estimated as
$L_{\rm ssc}(t) \propto t^{-2Y}$. 
The maximum frequency of the SSC emission $\nu_{\rm ssc}$
can be evaluated with Eq. (\ref{eq:gammaB}).
$\nu_{\rm ssc}
\sim\gamma^{2}\nu_{\rm syn}
\sim 1\times 10^{6}~ \gamma^{4}B~{\rm Hz}$ can be evaluated as  
\begin{eqnarray}
\nu_{\rm ssc}(t)\sim 1\times 10^{17}\gamma_{4}^{4}
B_{-5}
\left(\frac{t}{10^{7}~{\rm yr}}
\right)^{Y}
~{\rm Hz}
\end{eqnarray}
where we use Eq. (\ref{eq:gammaB})
and the typical value of  $B$ is set as
$B(t)= 10^{-5}~B_{-5}(t/10^{7}~{\rm yr})^{-Y}~{\rm G}$
based on Blandford (1990).
Thus, it is found that
non-thermal emissions from younger cocoon
reside in much lower energy band than MeV.


\section{Summary}

We have investigated the luminosity evolutions 
of AGN cocoons together with the dynamical 
evolution of expanding cocoon.
Below we summarize the main results of the present work.

\begin{itemize}
\item
We newly predict the bremsstrahlung 
emission peaked at MeV-$\gamma$ band as a result
of standard shock dissipation
of relativistic jets in AGNs. 
The temperatures of cocoon is 
governed only by the bulk Lorentz factor of the jet $\Gamma_{\rm j}$. 
The electron temperature $T_{e}$ 
relevant to observed emissions
is typically predicted in the range of  MeV
for $\Gamma_{\rm j}\sim 10$.
Constant temperatures of plasma in the cocoon
can be realized because of the continuous energy injection
by the jet with constant $\Gamma_{\rm j}$.
It should be emphasized that the constant behaviour
of AGN cocoon temperatures 
is different from the well known cases of  
gamma-ray bursts and supernovae.
In these sources, the temperatures decrease in time because the 
energy injection time scales are much shorter than their ages.
Since larger number densities of thermal electrons 
are predicted for younger cocoons, 
brighter thermal bremsstrahlung emission 
than that of older cocoon is naturally expected.

\item
Additionally,
non-thermal IC emissions from young cocoons
are also investigated. 
%
Importantly, in contrast to the case of MeV thermal emission,
the typical frequency of SSC and IC/CMB emissions
are predicted to be decreased for younger cocoon,
since the maximum Lorentz factor of 
relativistic electrons are decreased.
Therefore the typical  frequencies of IC from 
a younger cocoon are at much lower than MeV ranges.

\end{itemize}

\section*{Acknowledgments}

We are indebted to
M. C. Begelman, A. Celotti, F. Takahara and N. Isobe, J. Kataoka 
for valuable comments and discussions.
The anonymous referee is thanked for thoughtful comments
that improved the quality of the paper.
MK acknowledge the Grant-in-Aid  for Scientific Research of 
the Japanese Ministry of Education, Culture, Sports, Science and
Technology No. 14079025.  
HI acknowledge the Grant-in-Aid for the 21st Century COE Program
(Physics of Self-Organization Systems) and
a Grant for Special Research Projects at Waseda University.


\begin{thebibliography}{99}



\bibitem{BBR} Begelman M.~C., Blandford R.~D., Rees 
M.~J., 1984, Rev. Mod. Phys., 56, 255 


\bibitem{} Begelman, M. C. 1996, in 
Carilli C. L., Harris D. E., eds.,
Cygnus A---Study of a Radio Galaxy,  
Cambridge Univ. Press, Cambridge, p209

\bibitem{BC89} Begelman, M. C., \& Cioffi, D. F. 1989, ApJ, 345, L21(BC89)


\bibitem{} Blandford R.~D., Rees M.~J., 1974, MNRAS, 
169, 395 



\bibitem{B}
Blandford R. D., 1990, in Active Galactic Nuclei: 
Springer-Verlag, New York


\bibitem{bubble} Blanton E.~L., Sarazin C.~L., McNamara 
B.~R., Wise M.~W., 2001, ApJ, 558, L15 


\bibitem{bubble1} Boehringer H., Voges W., Fabian A.~C., 
Edge A.~C., Neumann D.~M., 1993, MNRAS, 264, L25 



\bibitem{Rlobe} Bridle, A. H., \& Perley, R. A. 1984, ARA\&A, 22, 319

\bibitem{} Celotti A., Fabian A.~C., 2004, MNRAS, 353, 523 


\bibitem{} Conway 
J.~E., 2002, NewAR, 46, 263 

\bibitem{Xlobe} Croston J.~H., Hardcastle M.~J., Harris 
D.~E., Belsole E., Birkinshaw M., Worrall D.~M., 2005, ApJ, 626, 733 



\bibitem{Drury} Drury L.~O., 1983, Rep. Prog. Phys., 46, 973 

\bibitem{} Dunn R.~J.~H., Fabian A.~C., 2004, 

MNRAS, 355, 862 

\bibitem{bubble} 
Fabian A.~C., et al., 2000, MNRAS, 318, L65 


\bibitem{Xlobe} Feigelson E.~D., Laurent-Muehleisen S.~A., 
Kollgaard R.~I., Fomalont E.~B., 1995, ApJ, 449, L149 

\bibitem{} Frederiksen J.~T., Hededal C.~B., 
Haugb{\o}lle T., Nordlund {\AA}., 2004, ApJ, 608, L13 




\bibitem{detector} Hasinger G., et al., 2001, A\&A, 365, L45 

\bibitem{Xlobe} Isobe, N., et al., 2002, ApJ, 580, L111

\bibitem{2002ApJ...565..163I} Iwamoto S., Takahara F., 2002, ApJ, 
565, 163 

\bibitem{Xcore} Kawaguchi T., Shimura T., Mineshige 
S., 2001, ApJ, 546, 966 

\bibitem{KK}
Kawakatu N., Kino M., 2006, MNRAS, 370, 1513 

\bibitem{Kino} Kino, M., Takahara, F., 2004, MNRAS, 349, 336

\bibitem{Kino} Kino, M., Kawakatu, N., 2005, MNRAS, 364, 659 

\bibitem{Xcore} Koratkar A., Blaes O., 1999, PASP, 111, 1 

\bibitem{} Leahy J.~P., Gizani N.~A.~B., 2001, 
ApJ, 555, 709 

\bibitem{b_hs} Liu, R., Pooley, G., \& Riley, J. M., 1992, MNRAS, 257, 545


\bibitem{M97} Mart\'{i}, M. A., et al., 1997, ApJ, 479, 151

\bibitem{Rlobe} Miley G., 1980, ARA\&A, 18, 165 

\bibitem{milne} 
Milne P.~A., Kroeger R.~A., Kurfess J.~D., The L.-S., 2002, NewAR, 46, 617 




\bibitem{} Nishikawa K.-I., Hardee P., Richardson G., 
Preece R., Sol H., Fishman G.~J., 2003, ApJ, 595, 555 


\bibitem{O'Dea98} O'Dea, C. P., \& Baum, S. A., 1997, AJ, 113, 148


\bibitem{Piran} Piran T., 1999, Phys. Rep., 314, 575 






\bibitem{negativeL} Readhead A.~C.~S., Taylor G.~B., Pearson 
T.~J., Wilkinson P.~N., 1996, ApJ, 460, 634 


\bibitem{pair} Reynolds C.~S., Fabian A.~C., Celotti A., 
Rees M.~J., 1996, MNRAS, 283, 873 



\bibitem{detector} 
Roques J.~P., et al., 2003, A\&A, 411, L91 

\bibitem{RL} 
Rybicki G. B., 
Lightman A. P., 1979,
Radiative Processes in Astrophysics, Wiley-Interscience, 
New York

\bibitem{S02} Scheck, L., et al., 2002, MNRAS, 331, 615, 2002 (S02)


\bibitem{cocoon} Scheuer P. A. G., 1974, MNRAS, 166, 513

\bibitem{b_hs} Scheuer, P. A. G., 1995, MNRAS, 277, 331

\bibitem{SM00} Sikora M., Madejski G., 2000, ApJ, 
534, 109 

\bibitem{} Shimada N., Hoshino M., 2000, ApJ, 
543, L67 


\bibitem{totani} Totani T., 1998, ApJ, 502, L13 

\bibitem{MeVdetector} Takahashi T., et al., 2004, NewAR, 48, 269 

\bibitem{vink} 
Vink J., Snellen I., Mack K.-H., Schilizzi R., 2006, MNRAS, 367, 928 



\bibitem{pair} 
Wardle J.~F.~C., Homan D.~C., Ojha R., Roberts D.~H., 1998, Nature, 395, 457 


\bibitem{} 
Weaver R., McCray R., Castor J., Shapiro P., Moore R., 1977, ApJ, 218, 377 






\end{thebibliography}
\end{document}